\makeatletter
\newcommand{\dontusepackage}[2][]{%
  \@namedef{ver@#2.sty}{9999/12/31}%
  \@namedef{opt@#2.sty}{#1}}
\makeatother
\dontusepackage{subfigure}
\documentclass[]{segabs}
\usepackage{amssymb,amsmath,amsbsy,bm}
\usepackage{fixltx2e} 
\IfFileExists{upquote.sty}{\usepackage{upquote}}{}
\usepackage[T1]{fontenc}
\usepackage[utf8]{inputenc}
\IfFileExists{microtype.sty}{\usepackage{microtype}}{}
\usepackage{natbib}
\usepackage{listings}
\usepackage{graphicx}
\makeatletter
\def\ScaleIfNeeded{%
  \ifdim\Gin@nat@width>\linewidth
    \linewidth
  \else
    \Gin@nat@width
  \fi
}
\makeatother
\let\Oldincludegraphics\includegraphics
{
 \catcode`\@=11\relax
 \gdef\includegraphics{\@ifnextchar[{\Oldincludegraphics}{\Oldincludegraphics[width=\ScaleIfNeeded]}}
}
\usepackage{caption}
\usepackage{float}
\usepackage{subfig}
\usepackage[unicode=true]{hyperref}
\hypersetup{breaklinks=true,
            bookmarks=true,
            pdfauthor={},
            pdftitle={Extended source imaging -- a unifying framework for seismic \& medical imaging},
            colorlinks=true,
            citecolor=black,
            urlcolor=blue,
            linkcolor=black,
            pdfborder={0 0 0}}
\def\argmin{\mathop{\rm arg\,min}}

\def\vector#1{\mathbf{#1}}

\def\tensor#1{\vector{#1}}

\usepackage{algorithm2e}

\title{Extended source imaging -- a unifying framework for seismic \& medical
imaging}
\author{Ziyi Yin, Rafael Orozco, Philipp Witte, Mathias Louboutin, Gabrio
Rizzuti, and Felix J. Herrmann\\School of Computational Science and
Engineering,\\Georgia Institute of Technology\\}
\date{}

\begin{document}
\maketitle

\begin{abstract}
We present three imaging modalities that live on the crossroads of
seismic and medical imaging. Through the lens of extended source
imaging, we can draw deep connections among the fields of wave-equation
based seismic and medical imaging, despite first appearances. From the
seismic perspective, we underline the importance to work with the
correct physics and spatially varying velocity fields. Medical imaging,
on the other hand, opens the possibility for new imaging modalities
where outside stimuli, such as laser or radar pulses, can not only be
used to identify endogenous optical or thermal contrasts but that these
sources can also be used to insonify the medium so that images of the
whole specimen can in principle be created.
\end{abstract}

\vspace*{-0.55cm}

\section{Introduction}\label{introduction}

\vspace*{-0.35cm}

The fields of seismic and medical imaging are both seeing major
developments in wave-equation based inversion
\citep{vanLeeuwen2013GJImlm, vanleeuwen2015IPpmp, Warner2016, Guasch809707, Huang2018};
in data acquisition with compressive sensing
\citep{herrmann2010GEOPrsg, liu2012compressed, sharan2018IEEEIUSspoa},
in large-scale solvers for convex optimization
\citep{zhang2014photoacoustic, esser2016tvr, peters2018pmf}, and in
machine learning
\citep{hauptmann2018approximate, herrmann2019NIPSliwcuc}. These
different techniques are aimed at improving acquisition efficiency
\citep{mosher2014increasing, kumar2017hrc}; image quality, via
hand-crafted \citep{esser2016tvr, peters2018pmf} or data-driven
regularization \citep{arridge2019solving}, and where possible to
quantify uncertainty \citep{siahkoohi2020EAGEdlb}. Obviously, there are
important commonalities between these two fields. They are both are
physics based, deal with data collection, intricate wave physics, and
both have a need to obtain high-fidelity and high-resolution images.
Despite these similarities, there are also important differences. For
instance, in medical imaging turn around times come at a premium while
in exploration seismology innovations in seismic data acquisition and
imaging are aimed at obtaining images in more and more complex
geological areas.

The fact that the goals and challenges in these fields are so different
in part explains why there are as of yet not been many examples of
successful synergetic developments in the fields of medical and seismic
imaging. In this talk, we try to overcome some of these hurdles by
studying three different examples, designed to exemplify connections
between seismic inversion with extended volume sources \citep{Huang2018}
and medical imaging modalities such as photoacoustic \citep{Xu2006} and
thermoacoustic imaging \citep{Ku2005}. Aside from presenting an unified
imaging framework for the different imaging modalities, we also propose
a novel imaging modality. Instead of locating optical or thermal
contrasts acoustically, we use the energy generated by these sources to
image spatial variations in the acoustic properties of the whole
specimen that can for example be used to locate calcium deposits
associated with cancer.

Our paper is organized as follows. First, we briefly describe our
general framework for wave-equation based imaging based on variable
projection. After introducing conventional active source imaging for an
unknown sourcetime signature, we shift our attention to extended volume
source imaging where the origin time and temporal source signature are
known but where the spatial location and radiation pattern are not. By
means of three examples, we demonstrate how these seemingly disconnected
formulations can be used to solve problems in photo/thermoacoustic and
seismic imaging.

\vspace*{-0.55cm}

\section{Wave-equation based imaging}\label{wave-equation-based-imaging}

\vspace*{-0.35cm}

Before discussing three different imaging modalities, we first present a
general framework deriving from the method of variable projection
\citep{aravkin2012IPNuisance}.

\vspace*{-0.15cm}

\subsection{Active source imaging}\label{active-source-imaging}

\vspace*{-0.15cm}

Inversion of Earth subsurface properties \citep{tarantola2005inverse}
has been a longstanding problem in exploration seismology. For active
sources, the following non-linear parameter estimation problem is
typically considered:
\begin{equation}
\min_{\mathbf{m},\mathbf{s}}f(\mathbf{m},\,\mathbf{s})=\frac{1}{2\sigma^2}\sum_{i=1}^{n_s}\|\mathbf{d}_i-F[\mathbf{m}]\ast_t\mathbf{q}^{\delta}_i[\mathbf{s}]\|_2^2 +\lambda R_t(\mathbf{s}).
\label{eq:FWI}
\end{equation}
 In this expression, the pair of unknowns $\{\mathbf{m},\, \mathbf{s}\}$
represent the discretized slowness squared and the temporal source
signature; $\sigma$ is the standard deviation of the noise of
multi-experiment data, $\{\mathbf{d}_i\}_{i=1}^{n_s}$, with $n_s$ the
number of source experiments (e.g.~shot records) generated by
$\{\mathbf{q}^{\delta}_i\}_{i=1}^{n_s}$ delta-like active sources with
known source positions, directivity patterns, and a typically unknown
temporal source signature $\mathbf{s}$. To underline linearity in the
source, and the convolutional relationship between the source time
signature and the receiver restricted Green's function
$F[\mathbf{m}]=\mathbf{P}_r\mathbf{A}^{-1}[\mathbf{m}]\cdot$, with
$\mathbf{A}^{-1}[\mathbf{m}]\cdot$ the inverse of the discretized wave
equation and $\mathbf{P}_r$ the receiver restriction operator, we
introduced the symbol $\ast_t$ to represent convolutions over time. To
prevent overfitting of the source (see \citet{yang2020timedomain} for
detail), we added the $\lambda$-weighted penalty term $R_t(\mathbf{s})$.

In active source seismic exploration, the sources
$\{\mathbf{q}^{\delta}_i\}_{i=1}^{n_s}$ are assumed spatially impulsive
and considered as discretizations of the outer product between spatial
Dirac distributions, centered at the source locations
$\delta(\mathbf{x}-\mathbf{x}^s_i)s(t)^\top$ with $\mathbf{x}=(x,\,z)$
the spatial coordinates in 2D, and a single unknown but causal source
time function $s(t)=0,\, t\leq 0$ represented by the vector
$\mathbf{s}$.

To solve the above optimization problem, we rely on the technique of
variable projection \citep{aravkin2012IPNuisance} where the following
reduced objective is minimized:
\begin{equation}
\min_{\mathbf{m}} \bar{f}(\mathbf{m})=f(\mathbf{m},\,\bar{\mathbf{s}}(\mathbf{m}))\quad\text{where}\quad\mathbf{\bar{s}}=\argmin_{\mathbf{s}}f(\mathbf{m},\,\mathbf{s}).
\label{eq:reduced-fwi}
\end{equation}
 We obtain the reduced objective $\bar{f}(\mathbf{m})$ in
Equation~\ref{eq:reduced-fwi} by substituting the temporal source
function that minimizes the objective $f(\mathbf{m},\mathbf{s})$, for
fixed $\mathbf{m}$, into the objective of Equation~\ref{eq:FWI}. To
arrive at this approximate first-order accurate formulation, we made use
of the fact that $\nabla_{\mathbf{s}}f(\mathbf{m},\,\bar{s})=0$ at the
optimal point, which for a fixed $\mathbf{m}$ minimizes the FWI
objective.

This solution method proceeds by updating the model vector $\vector{m}$
by computing the gradient of the reduced objective that has the
following form:
\begin{equation}
    \mathbf{g}=\nabla \bar{f}(\mathbf{m})=\frac{1}{\sigma^2}\sum_{i=1}^{n_s}\nabla\mathbf{F}_i^{\top}\left(\mathbf{d}_i-F[\mathbf{m}]\ast_t\mathbf{q}_i[\mathbf{\bar{s}}]\right)
\label{eq:gradient-fwi}
\end{equation}
 with $\nabla\mathbf{F}_i$ the Jacobian for the $i\text{th}$ source. In
this expression, the gradient inherited the sum structure of the above
FWI objective, where each term in the sum corresponds to contributions
from different active source experiments with the source locations
considered to be known. Because the temporal source signature is related
to the data through a simple convolution, its inversion
(cf.~Equation~\ref{eq:reduced-fwi}) is relatively straightforward to
solve. As recently shown by \citet{yang2020timedomain}, this can be done
on-the-fly in the time-domain with code automatically generated by
\href{https://github.com/devitocodes/devito}{Devito}
\citep{devito-compiler, devito-api} and called by the Julia Devito
Inversion framework
\citep[\href{https://github.com/slimgroup/JUDI.jl}{JUDI},][]{witteJUDI2019}.
The combination of these two frameworks allows us to scale to 3D and to
include more complex wave physics.

\vspace*{-0.15cm}

\subsection{Extended Volume Source
Imaging}\label{extended-volume-source-imaging}

\vspace*{-0.15cm}

While inversions based on active source experiments have their
counterparts in ultrasonic medical imaging, this is not the only imaging
modality available to the medical practitioner. Contrary to more or less
exclusive use of active acoustic sources, such as dynamite, air guns, or
(marine) vibrators, the field of medical imaging successfully developed
alternative modalities where in situ ultrasonic acoustic sources are
triggered by synchronized external stimuli such laser light pulses, as
in photoacoustic imaging \citep{Xu2006}; pulses of radio waves, as in
thermoacoustic imaging \citep{Ku2005}; or even ultrasound waves
themselves, as in acoustic cavitation \citep{peshkovsky2010acoustic}. In
all cases, the external source triggers ultra-sonic acoustic sources
deep inside the medium of interest and this offers unique possibilities
that one normally would not have in conventional active source imaging.
Contrary to the ``seismic'' active source setting, the spatial position,
and even the shape, of the stimuli-induced secondary sources are not
known. However, because electro-magnetic waves travel much faster than
acoustic waves, the firing times of these induced sources are known
except for the case of acoustic cavitation where the outside stimulus
itself travels with the speed of sound.

In situations where the firing (origin) times and pulse shapes are known
but where the spatial distribution of the secondary sources are unknown,
our (non-)linear parameter estimation problem becomes:
\begin{equation}
\min_{\mathbf{m},\,\mathbf{u}}{f}(\mathbf{m},\,\mathbf{u})=\frac{1}{\sigma^2}\sum_{i=1}^{n_s}\|\mathbf{d}_i-{F}[\mathbf{m}]\ast_t\mathbf{{q}}^e[\mathbf{u}_i]\|_2^2+\lambda R_x(\mathbf{u}_i)
\label{eq:opto}
\end{equation}
 where the extended sources
${\mathbf{q}}^{e}[\mathbf{u}_i],\, i=1\cdots n_s$ are now given by a
discretization of the outer product $u_i(\mathbf{x})s(t)^\top$ between
an extended unknown spatial source distribution $u(\mathbf{x})$ and a
known temporal causal source function $s(t)$ firing at $t=0$. As before,
we include a penalty term, $R_x(\mathbf{u}_i)$, to regularize inversion
of the extended sources. For simplicity, we drop the subscript $i$ for
the optimization variable.

Like with regular FWI, we can derive a reduced objective for the
extended source problem via
\begin{equation}
\min_{\mathbf{m}} \bar{f}(\mathbf{m})=f(\mathbf{m},\,\bar{\mathbf{u}}(\mathbf{m}))\quad\text{where}\quad\mathbf{\bar{u}}=\argmin_{\mathbf{u}}f(\mathbf{m},\,\mathbf{u})
\label{eq:reduced-opto}
\end{equation}
 and proceed by calculating the gradient via
\begin{equation}
    \mathbf{g}=\nabla \bar{f}(\mathbf{m})=\frac{1}{\sigma^2}\sum_{i=1}^{n_s}\nabla\mathbf{F}^{\top}\left(\mathbf{d}_i-F[\mathbf{m}]\ast_t\mathbf{q}[\mathbf{\bar{u}}_i]\right)
\label{eq:gradient-opto}
\end{equation}
 where the $\{\mathbf{u}_i\}_{i=1}^{n_s}$ are computed by solving
Equation~\ref{eq:reduced-opto} for each source experiment separately.

While the reduced formulations in Equations~\ref{eq:reduced-fwi}
and~\ref{eq:reduced-opto} are conceptually similar, inverting the
source-time function is, because of the convolutional structure, simpler
and does not need evaluations of $F[\mathbf{m}]$ and its adjoint to
project out the temporal source. This is not the case for the volume
extended source, whose estimation is expensive and requires
regularization via the spatial penalty term $R_x(\mathbf{u})$ or via
constraints. In the next section, we show how this formulation serves as
the basis of an integrated imaging framework for seismic and medical
imaging.

\vspace*{-0.55cm}

\section{Connecting the dots}\label{connecting-the-dots}

\vspace*{-0.35cm}

We will now show how Equations~\ref{eq:opto} and~\ref{eq:gradient-opto}
can be used to solve seemingly different problems in medical and seismic
imaging. Constraints on the extended source are implemented with the
software
\href{https://github.com/slimgroup/SetIntersectionProjection.jl}{SetIntersectionProjection}
(\citet{peters2018pmf}).

\vspace*{-0.15cm}

\subsection{Case I---Photoacoustic imaging w/
constraints}\label{case-iphotoacoustic-imaging-w-constraints}

\vspace*{-0.15cm}

\begin{figure}
\centering
\subfloat[\label{fig:case-I-setup}]{\includegraphics[width=0.500\hsize]{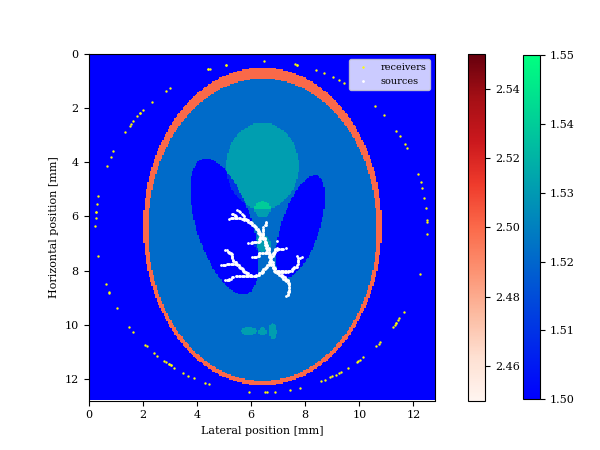}}
\subfloat[\label{fig:case-I-adjoint}]{\includegraphics[width=0.500\hsize]{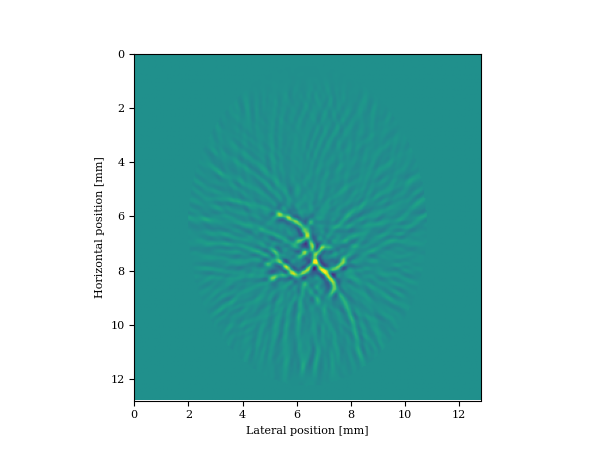}}
\\
\subfloat[\label{fig:case-I-inversion}]{\includegraphics[width=0.500\hsize]{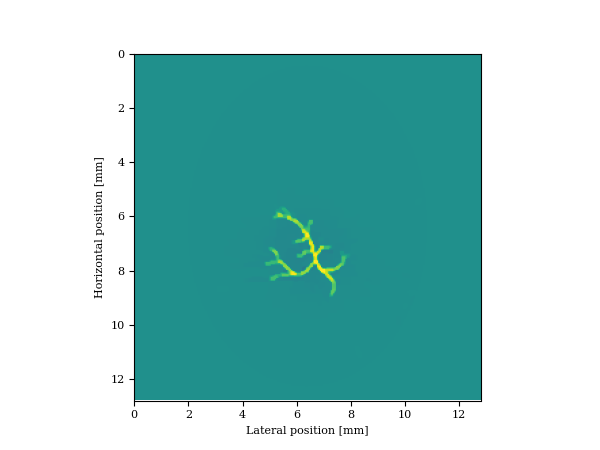}}
\subfloat[\label{fig:case-I-water}]{\includegraphics[width=0.500\hsize]{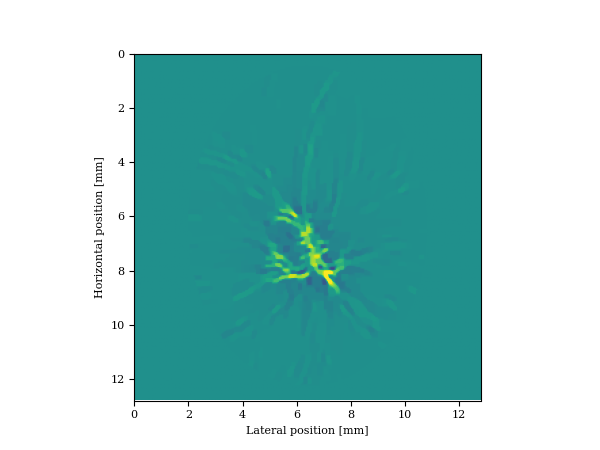}}
\caption{Compressive imaging in miniaturized Shepp Logan Phantom.
\emph{(a)} Experimental setup with 100 randomly receivers (yellow
$\cdot$'s) and blood vessels of interest in white. \emph{(b)}
Reconstructed image using back propagation. \emph{(c)} The same using
inversion of Equation~\ref{eq:opto-constrained} with the TV-constraint.
\emph{(d)} The same but imaged with constant water
velocity.}\label{fig:case-I}
\end{figure}

Perhaps the most straightforward application of Equation~\ref{eq:opto}
is in photoacoustic imaging where an unknown distribution of endogenous
optical or thermal contrast sources, the object of interest, are
stimulated by laser beams or radio waves. In most applications of this
imaging modality, the constant or smoothly varying acoustic velocity
model is assumed to be known and the ``exploding reflector'' induced by
the laser pulse is the unknown. While good results via back propagation
of the observed wavefields are possible, these results typically rely on
high fidelity data collected at high spatial sampling rates
\citep{cox2007k}, which puts pressure on the acquisition system where
large numbers of channels come at a premium. We show that by adding a
hand-crafted constraint (total-variation norm in this case, see also
\citet{zhang2014photoacoustic}; \citet{sharan2018IEEEIUSspoa}) to
Equation~\ref{eq:opto}, we can remove subsampling related artifacts by
solving for a given smooth background velocity model $\mathbf{m}_0$
\begin{equation}
\min_{\mathbf{u}}\frac{1}{\sigma^2}\|\mathbf{d}-{F}[\mathbf{m}]\ast_t\mathbf{{q}}^e[\mathbf{u}]\|_2^2\quad\text{s.t.}\quad \|\mathbf{u}\|_{TV}\leq \tau.
\label{eq:opto-constrained}
\end{equation}
 In this equation, we added the total-variation norm as a handcrafted
constraint using the approach of \citet{peters2018pmf}. To evaluate the
performance of this approach, we consider a thermoacoustic imaging
example of a miniaturized Shepp Logan phantom, which we converted to
acoustic wavespeeds \citep{clement2013}. This imaging problem that
operated at $5\,\mathrm{MHz}$ is challenging because of the skull, which
has a high velocity of $2.5\,\mathrm{km/s}$. Assuming that radio waves
can penetrate the skull, a thermoacoustic response can be triggered
emanating from the white-color blood vessels (see
Figure~\ref{fig:case-I-setup}). To further complicate things, we
fivefold randomly subsampled the receivers and compare the
reconstruction via back propagation with solving
Equation~\ref{eq:opto-constrained}. Data is generated in the true
velocity model while the inversion is carried out in a smoothed
kinematically correct velocity model. The results are summarized in
Figure~\ref{fig:case-I} from which we can draw the following
conclusions: \emph{(i)} as predicted by compressive sensing, inversions
with the TV-norm constraint are robust w.r.t. randomized receiver
subsampling compared to imaging via back propagation
(cf.~Figures~\ref{fig:case-I-adjoint} \&~\ref{fig:case-I-inversion}) and
\emph{(ii)} carrying out the imaging in a spatially varying velocity
model (not constant water speed) is essential in order to image through
the skull (cf.~Figures~\ref{fig:case-I-inversion}
\&~\ref{fig:case-I-water}).

\vspace*{-0.15cm}

\subsection{Case II --- Seismic imaging w/ extended
sources}\label{case-ii-seismic-imaging-w-extended-sources}

\vspace*{-0.15cm}

Even though Equation~\ref{eq:opto} seems outside the realm of regular
exploration seismology, there is a direct relation between this equation
and recently proposed extensions of FWI designed to mitigate the effects
of cycle skipping---i.e., the existence of parasitic local minima. To
mitigate the possible adverse effects of these local minima, several
``explain away'' approaches have been proposed where additional
slack/latent variables are introduced that give the problem more freedom
to fit observed data even though the starting model for the velocities
is wrong. Without these extra variables, the gradient has a tendency to
point in the wrong direction sending FWI on a road of failure.

Extended formulations, on the other hand, may overcome this problem, in
certain situations, by fitting the data, by minimizing over the slack
variable first, followed by computing the gradient of the reduced
objective as described above (cf.~Equations~\ref{eq:opto}
and~\ref{eq:reduced-opto}). Examples of this type of approach include
Adaptive Waveform Inversion \citep[AWI,][]{Warner2016}, where
trace-by-trace Wiener filters are introduced that match the observed and
simulated traces, followed by a model update designed to turn these
filters into zero-phase spikes; Wavefield Reconstruction Inversion
\citep[WRI,][]{vanLeeuwen2013GJImlm, vanleeuwen2015IPpmp}, where a
data-augmented wave equation is solved that matches the physics as well
as the observed data, followed by taking a gradient step that updates
the velocity model such that the wave equation itself holds (i.e., by
focusing the augmented wavefield onto the sources); and extended
waveform inversion with volumetric sources \citep{Huang2018}, where
source extensions are used to match the observed data, followed by
computing the gradient of the reduced objective. Equation~\ref{eq:opto}
is an instance of this latest approach where a source focusing penalty
term is added as regularization. In this case we have,
$R(\mathbf{u})=\|\mathbf{W}\mathbf{u}\|_2^2$ where $\mathbf{W}$ is an
annihilator constructed in such a way that the extended source is in its
null space when focussed---i.e., $R(\delta(\mathbf{x}-\mathbf{x}_s))=0$.
This can be achieved by setting
$\mathbf{W}\mathbf{u}=|\mathbf{x}-\mathbf{x}_s|\mathbf{u}$
\citep{Huang2018}.

\begin{figure}
\centering
\subfloat[\label{fig:case-II-setup}]{\includegraphics[width=0.800\hsize]{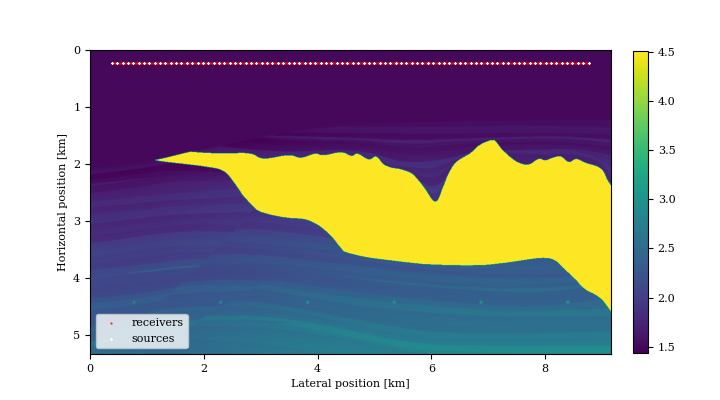}}
\\
\subfloat[\label{fig:case-II-rtm}]{\includegraphics[width=0.800\hsize]{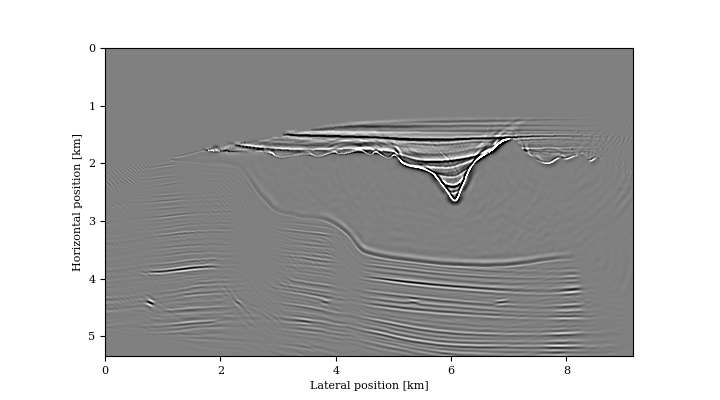}}
\\
\subfloat[\label{fig:case-II-inversion}]{\includegraphics[width=0.800\hsize]{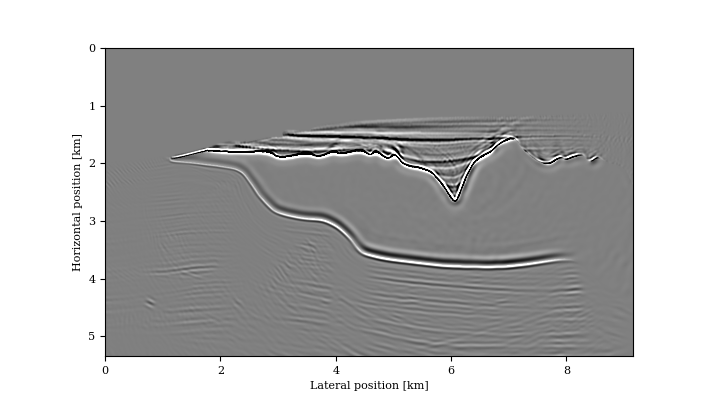}}
\caption{Comparison of extended source imaging and conventional
reverse-time migration. \emph{(a)} Experimental setup with $90$ sources
and $500$ receivers. \emph{(b)} Result obtained by conventional
migration with the inverse-scattering imaging condition. \emph{(c)}
Result obtained by solving for the extended sources first by minimizing
Equation~\ref{eq:opto} for $\mathbf{u}$, followed by computing the
gradient of the reduced objective (Equation~\ref{eq:gradient-opto}) with
the inverse-scattering condition. While there are some differences in
the frequency content and amplitudes, the overall image quality is
similar albeit that the bottom salt is improved by extended source
imaging.}\label{fig:case-II}
\end{figure}

While the above approach has been used successfully to mitigate some of
the effects of local minima in FWI, we show that this approach can also
be used to migrate by solving for $\mathbf{\bar{u}}$ first, by running
LSQR on Equation~\ref{eq:opto}, followed by computing the gradient of
the reduced objective with the inverse-scattering imaging condition
\citep{witte2018cls}, which is designed to reduce tomographic artifacts.
As we can see in Figure~\ref{fig:case-II}, this approach is indeed
capable of producing high-fidelity images in complex models without the
need to know the exact location and the directivity pattern of the
sources. The interesting aspect of this two-stage approach is that we
can image as long as we know the firing times of the different source
experiments, which opens a new perspective on medical imaging.

\vspace*{-0.15cm}

\subsection{Case III --- Photoextended
imaging}\label{case-iii-photoextended-imaging}

\vspace*{-0.15cm}

Contrary to seismic imaging, the field of medical imaging comprises of a
wide range of different imaging modalities where different external
stimuli are used to create an image with induced ultrasonic waves.
Photoacoustic imaging is a good example of such an approach where laser
light induces ultrasonic contrast sources, which can be imaged as we
described under Case I. However, we can take this imaging modality a
step further by using the methodology described under Case II. Contrary
to conventional photoacoustic imaging, we use these contrasts as
``active'' sources insonifying the specimen as a whole.

\begin{figure}
\centering
\subfloat[\label{fig:case-III-setup}]{\includegraphics[width=0.470\hsize]{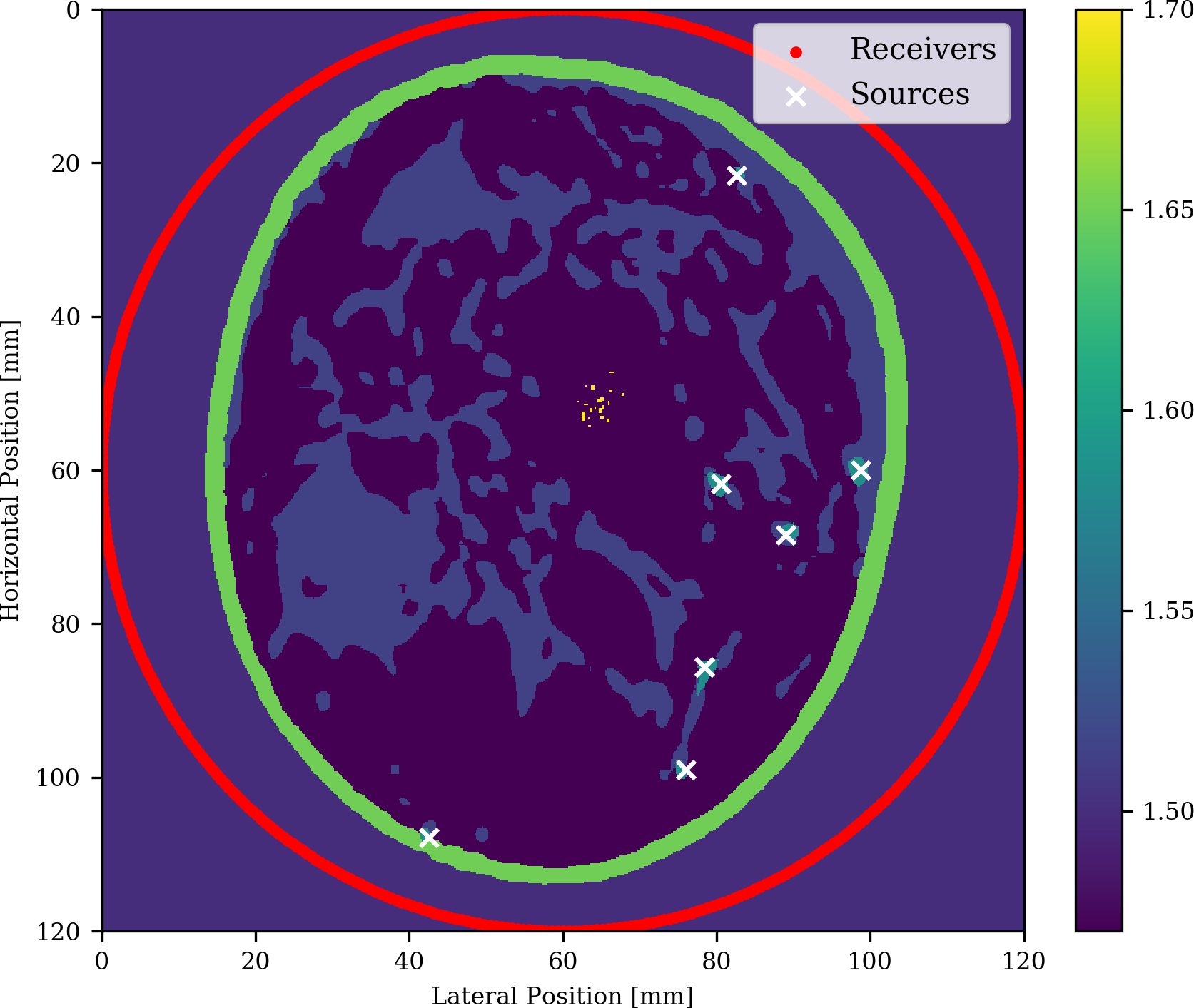}}
\\
\subfloat[\label{fig:case-III-single}]{\includegraphics[width=0.470\hsize]{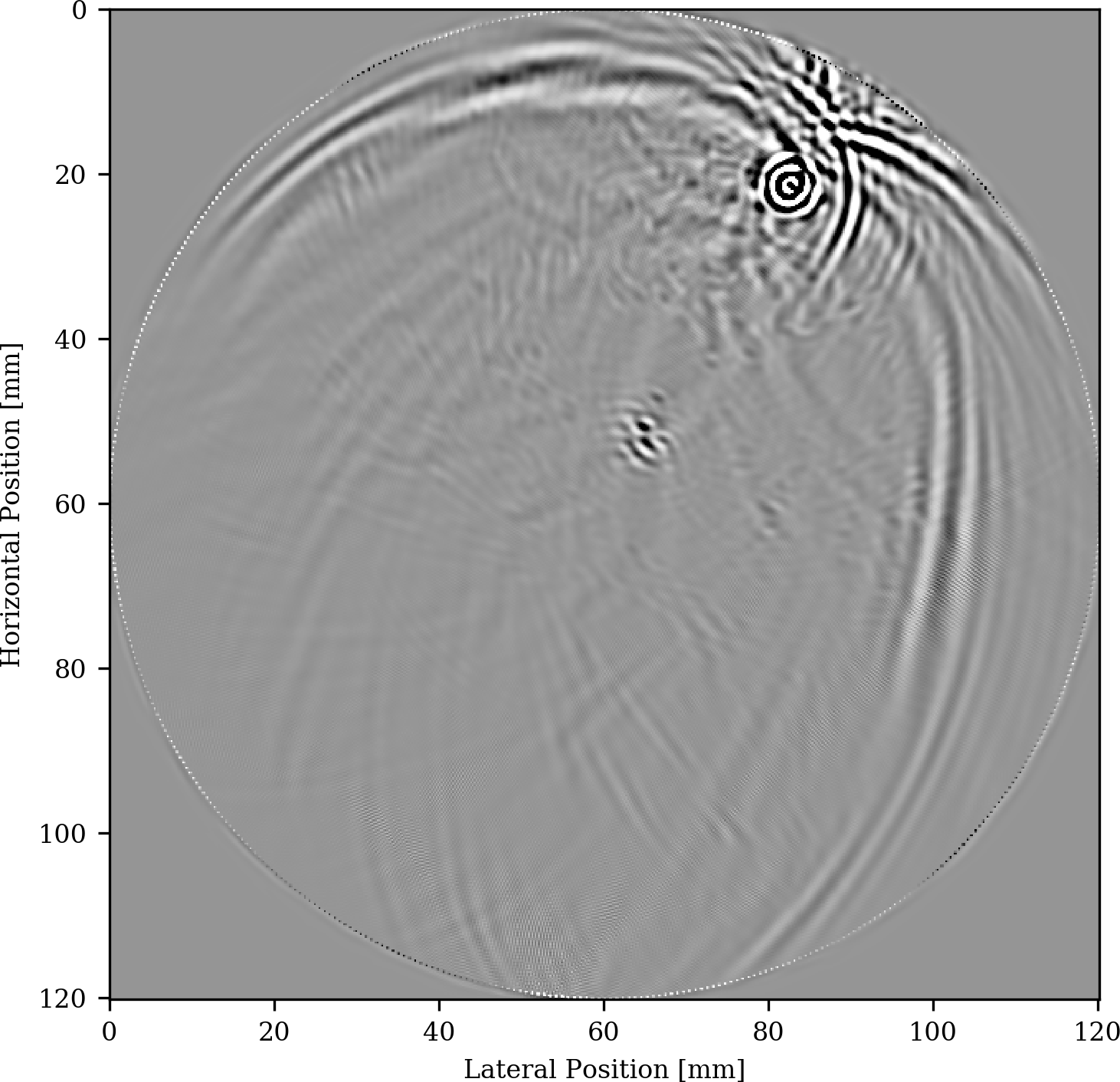}}
\\
\subfloat[\label{fig:case-III-stack}]{\includegraphics[width=0.470\hsize]{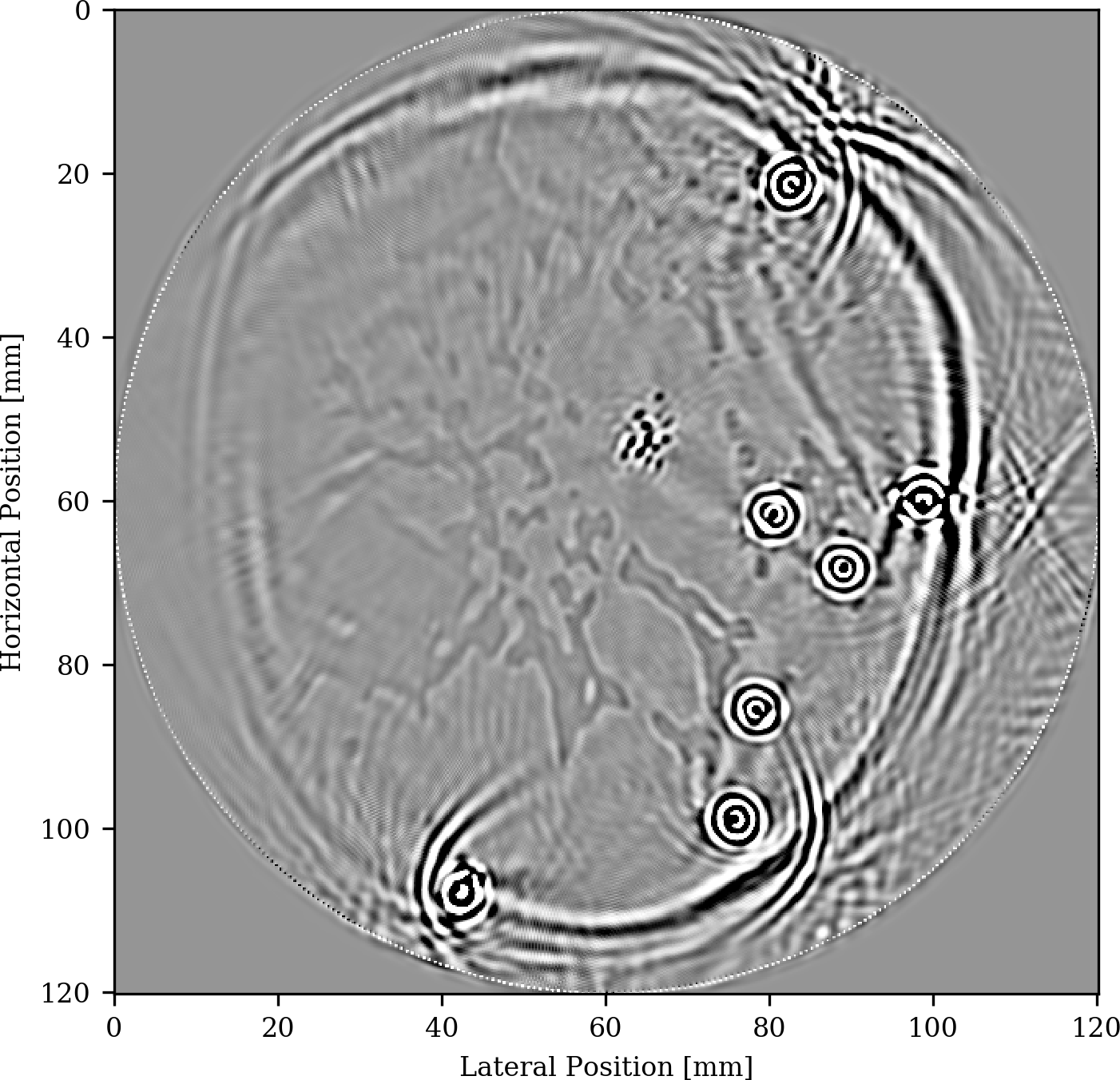}}
\caption{Example of photoextended imaging where blood vessels act as
sources for a breast with speckle microcalcifications present.
\emph{(a)} Experimental setup with $7$ sources denoted by white
$\times$'s, $512$ receivers denoted by red $\cdot$'s, and actual
wavespeed in the color image. \emph{(b)} Image obtained for a single
extended source in a smoothly varying background velocity model.
\emph{(c)} Stacked image over $7$ extended sources places throughout the
breast. Aside from clearly observing the cluster of microcalcifications,
we also obtain an image of fat and fibroglandular
tissue.}\label{fig:case-III}
\end{figure}

For this purpose, let us consider a 2D example of breast imaging where
we are interested in finding microcalcifications (calcium oxalate) that
could be indicative of breast cancer. To create an image, we assume that
we are able to carry out independent photo-source experiments where
ultrasonic ``point sources'' are triggered with laser light. We assume
that these sources are located in blood vessels. We repeat this multiple
times. Since we do not know where the blood vessels are, we do not apply
focussing but instead we solve for each source separately using $5$
iterations of LSQR, followed by computing the gradients, as in
Equation~\ref{eq:gradient-opto}, which we stack into a single image. The
setup for our experiment is depicted in Figure~\ref{fig:case-III-setup}
and includes $7$ different opto-induced sources placed in blood vessels
and denoted by the white crosses and densely sampled receivers placed in
a circular fashion around the breast. As we can see from
Figure~\ref{fig:case-III-stack}, we obtain a clear image not only of the
microcalcifications but also of the fat and fibroglandular tissue.

\vspace*{-0.55cm}

\section{Observations}\label{observations}

\vspace*{-0.35cm}

We believe that these experiments demonstrate that imaging algorithms
originally motivated by specific challenges encountered in seismic
problems can be applied in some interesting medical imaging scenarios.
We make this assertion under the assumption that we are in the correct
physical regime and that we have access to adequate computational
resources. If this is indeed the case, there are exciting possibilities
because wave-equation based inversion, in tandem with regularization
with contraints and deep priors, opens the way towards high-fidelity
high-resolution images including uncertainty quantification.

\vspace*{-0.15cm}

\subsection{Related materials}\label{related-materials}

In order to facilitate the reproducibility of the results herein
discussed, a Julia implementation of this work is made available on the
\href{https://slim.gatech.edu}{SLIM} GitHub page
\url{https://github.com/slimgroup/Software.SEG2020}.

\vspace*{-0.15cm}

\subsection{Acknowledgements}\label{acknowledgements}

\vspace*{-0.15cm}

Felix J. Herrmann would like to thank the Georgia Institute of
Technology and the Georgia Research Alliance for their financial
support.

\bibliography{paper}

\end{document}